\documentclass[%
 reprint,
 amsmath,amssymb,
 aps,
]{revtex4-2}

\usepackage{graphicx}
\usepackage{dcolumn}
\usepackage{bm}

\begin{document}

\preprint{APS/123-QED}

\title{Conjecture of an information-response inequality}

\author{Andrea Auconi}
\affiliation{%
 Ca’ Foscari University of Venice, DSMN, via Torino 155, 30172 Mestre (Venice), Italy
}%

\date{\today}

\begin{abstract}
The invariant response was defined from a formulation of the fluctuation-response theorem in the space of probability distributions. An inequality is here conjectured which sets the mutual information as an upper bound to the invariant response, and its large information limit is proven. Applications to the thermodynamics of feedback control and to estimation theory are discussed.
\end{abstract}

\maketitle



The response is the statistics of the propagation of small perturbations in a stochastic dynamical system.
The fluctuation-response theorem establishes that the response can be predicted from properties of the unperturbed dynamics \cite{marconi2008fluctuation,dechant2020fluctuation}.
From a formulation of the fluctuation-response theorem in the space of probability distributions we defined a response function that is invariant under invertible transformations \cite{auconi2021fluctuation}.
This invariant response is an abstraction from the response function of single observables to that of the whole probability distribution, in analogy to how mutual information generalizes linear correlations \cite{cover1999elements,dembo1991information,rioul2010information}. 


The mutual information is an invariant measure of the strength of probabilistic dependencies, and its conditional version is used to quantify directed information flow in thermodynamics \cite{ito2013information,horowitz2014thermodynamics,auconi2019information}.
For the case of linear interactive dynamics we found that an exact relation holds between invariant response and mutual information \cite{auconi2021fluctuation}.

The invariant response involves derivatives of the probability distribution, therefore it has a local character which makes it qualitatively different from the mutual information. Driven by mathematical curiosity, instead of the full dynamic problem here we consider an input-output setting where the connection between these two probability functionals is more easily studied.

In this Letter, an inequality is conjectured which sets the mutual information as an upper bound to the invariant response. A proof of the large information limit of this conjecture is given by mapping it to Stam's isoperimetric inequality \cite{dembo1991information}. From this conjecture we could derive as special cases two known results in thermodynamics \cite{sagawa2010generalized} and estimation theory \cite{barnes2021fisher}.

\paragraph*{Mutual information.}
Consider two smooth probability densities $p(x)$ and $q(x)$, and denote by $\langle f \rangle$ the expectation of a function $f(x)$ with respect to $p(x)$.
The Kullback-Leibler (KL) divergence \cite{amari2016information} from $q$ to $p$ is defined as $D\left[ p||q \right]\equiv \left\langle \ln (p/q) \right\rangle$, it is nonnegative and invariant under invertible transformations $x\rightarrow x'$.
Shannon's mutual information \cite{cover1999elements} is defined as the KL divergence from independence of the bivariate joint probability $p(x,y)$, namely $I \equiv D\left[ p(x,y)|| p(x) p(y) \right] $.


\paragraph*{Invariant response.}
Let us consider an input-output setting common in communication theory \cite{cover1999elements} where the marginal $p(y)=\int dx \, p(x) \, p(y|x)$ is interpreted as an output density uniquely determined by the input density $p(x)$ by passing it through the fixed channel $p(y|x)$.
In this framework let us introduce a simple perturbation of the input density, $q(x)=p(x-\epsilon)$, which corresponds to a translation on the right by a small quantity $\epsilon$. We quantify the magnitude of this perturbation by the KL divergence from $p$ to $q$, namely $D\left[ q(x) || p(x) \right] $, and it can be interpreted as a cost of the perturbation \cite{auconi2021fluctuation}.
The perturbation propagates to the output density as $q(y)=\int dx \, q(x) \, p(y|x)$, and accordingly we quantify the response with the KL divergence $D\left[ q(y) || p(y) \right]$. We define the invariant response $\gamma_{x\rightarrow y}$ as the small perturbation limit $\epsilon\rightarrow 0$ of the response-perturbation divergences ratio $D\left[ q(y) || p(y) \right] / D\left[ q(x) || p(x) \right]$, which gives
\begin{equation}\label{definition}
   \left\langle \left( \partial_x \ln p(x) \right)^2 \right\rangle \gamma_{x\rightarrow y} 
   =  \left\langle \left\langle \partial_x \ln p(x) \big{|} y \right\rangle ^2 \right\rangle,
\end{equation}
that is a form of the fluctuation-response theorem in the space of probability distributions \cite{auconi2021fluctuation}.

The partial derivative operator $\partial_x$ coming from the small perturbation expansion and typical of the Fisher information \cite{dembo1991information}, makes the invariant response $\gamma_{x\rightarrow y}$ qualitatively different from the mutual information $I$.
Also let us note that while the mutual information $I$ is symmetric, in general the invariant response is not, $\gamma_{x\rightarrow y}\neq \gamma_{y\rightarrow x}$.

Applying Jensen's inequality on the right hand side of Eq. \eqref{definition} gives $0 \leq \gamma_{x\rightarrow y} \leq 1$, which is a type of data processing inequality \cite{dembo1991information,rioul2010information} as we discuss below.
From Cauchy-Schwarz inequality instead we obtain a bound in terms of the $\chi^2$ divergence \cite{amari2016information}, namely $0 \leq \gamma_{x\rightarrow y} \leq \left\langle p(y|x)/p(y) \right\rangle -1$, which does not seem to have a simple physical interpretation.

\paragraph*{The conjecture.}
We conjecture the following information-response inequality
\begin{equation}\label{conjecture}
    \gamma_{x\rightarrow y} \leq 1 - e^{-2 I},
\end{equation}
with equality for bivariate Gaussians. By symmetry the same inequality should hold for $\gamma_{y\rightarrow x}$ if we model with respect to the opposite channel $p(x|y)$.

The equality case in Eq. \eqref{conjecture} can be immediately verified by taking a bivariate Gaussian for $p(x,y)$.
Below we prove that the large information limit of the conjecture is correct by using Stam's isoperimetric inequality \cite{dembo1991information}. We also show that the Gaussian family is a stationary point of the Lagrange multipliers problem for maximising the functional $\gamma_{x\rightarrow y}$ with fixed mutual information $I$ and probability normalization. Moreover, we numerically verified the conjecture on a particular class of random densities realizations, see Supplementary Materials (SM).

\paragraph*{Refined data processing inequality.}
Consider a Gaussian noise source $z\sim \mathcal{N}(0,\epsilon)(z) \equiv [2\pi\epsilon]^{-\frac{1}{2}} \exp[ -\frac{z^2}{2\epsilon} ]$, and pass it through the channel $p(y|x)$ by perturbing the input $\widetilde{x} = x+z$, diagrammatically $z\rightarrow \widetilde{x}\rightarrow \widetilde{y}$. In the limit of small variance $\epsilon\rightarrow 0$ we find that the data processing ratio converges to the invariant response, $\gamma_{x\rightarrow y} = \lim_{\epsilon \rightarrow 0} \left[ I(z,\widetilde{y}) / I(z,\widetilde{x}) \right]$,
in a type of De Bruijn's identity \cite{dembo1991information,rioul2010information}, see SM.
Since mutual information is nonnegative, and the channel $p(y|x)$ should satisfy the data processing inequality \cite{dembo1991information,rioul2010information}, $I(z,\widetilde{y}) \leq I(z,\widetilde{x})$, then we obtain again $0 \leq \gamma_{x\rightarrow y} \leq 1$ but from information-theoretic arguments only.
We see that the conjecture of Eq. \eqref{conjecture} can be interpreted as a refined data processing inequality based on the channel mutual information, namely $\lim_{\epsilon \rightarrow 0} \left[ I(z,\widetilde{y}) / I(z,\widetilde{x}) \right] \leq 1-\exp \left[ - 2 I(\widetilde{x},\widetilde{y})\big{|}_{\epsilon=0} \right]$.

\paragraph*{Application in thermodynamics.}

Consider a Brownian particle with stochastic dynamics $dx=-\partial_x U(x) \,dt + \sigma \,dW$, in a harmonic potential $U(x)=x^2 /2$ so that the stationary density is $p(x)=\mathcal{N}(0,\sigma^2/2)$.
A shift of the potential to $V(x,\lambda)=U(x-\lambda)$ performs a work on the particle $w(x,\lambda)=V(x,\lambda)-U(x)=\lambda^2/2-\lambda x$.
Consider a measurement of $x$ according to a generic non-Gaussian density $p(y|x)$. The expected work after a measurement $y$ for a fixed shift $\lambda$ is
$ \left\langle w \big{|} y,\lambda \right\rangle = \lambda^2/2 - \lambda \left\langle x \big{|} y \right\rangle$.
To efficiently use the information from the measurement we optimize $\lambda$ to maximize the extracted work obtaining simply $\lambda(y) = \left\langle x \big{|} y \right\rangle$ so that $ \left\langle w^* \big{|} y \right\rangle \equiv \left\langle w \big{|} y, \lambda(y) \right\rangle = -\left\langle x \big{|} y \right\rangle^2 /2 $. Averaging over many realizations we obtain $ \left\langle w^* \right\rangle = -\left\langle \left\langle x \big{|} y \right\rangle^2 \right\rangle /2 $.
The bivariate density in this problem is $p(x,y)= \mathcal{N}(0,\sigma^2/2)(x) \,p(y|x)$, so that the invariant response is $\gamma_{x\rightarrow y} = \left\langle\left\langle x \big{|} y \right\rangle^2 \right\rangle / \left\langle x^2 \right\rangle = - 4 \left\langle w^* \right\rangle / \sigma^2 $. Then by using the conjecture of Eq. \eqref{conjecture} we get
\begin{equation}
    \left\langle w^* \right\rangle \geq - \frac{\sigma^2}{4} \left( 1 - e^{-2 I} \right) ,
\end{equation}
which is a tighter bound compared to $-\frac{\sigma^2}{2}  I$, the one coming from the feedback fluctuation theorem \cite{sagawa2010generalized}.
In particular, the bound derived from the conjecture limits the optimal work extraction to the particle energy $\frac{\sigma^2}{4}$.

\paragraph*{Application in estimation theory.}
The Fisher information is a central quantity in estimation theory as it lower bounds the error in estimating model parameters from data via the Cramer-Rao bound \cite{dembo1991information,rioul2010information}.

A mean-zero random variable $x$ is said to be sub-Gaussian with parameter $b$ if $\left\langle e^{t x} \right\rangle \leq e^{t^2 b^2 /2}$ holds for all $t \in \mathbb{R}$.
A recent result in estimation theory \cite{barnes2021fisher} demonstrates a special inequality for the case when $\partial_x \ln p(x)$ is sub-Gaussian, that in our formalism reads
$\left\langle \left\langle \partial_x \ln p(x) \big{|} y \right\rangle ^2 \right\rangle
\leq 2 b^2 I$. We can derive this result from our conjecture by noting that $\left\langle \left( \partial_x \ln p(x) \right)^2 \right\rangle
\leq b^2$, see \cite{rivasplata2012subgaussian}, and using the convexity $e^x\geq 1+x$.

\paragraph*{Example of vanishing response.}
The bivariate Gaussian family is the equality case in the conjecture. We here discuss the opposite case of finite mutual information and vanishing invariant response. This is achieved, for example using a Gaussian channel $p(y|x)=\mathcal{N}(x,1)$, with a fast oscillatory input distribution like $p(x)=\cos^2(\alpha x) \, e^{-x^2} /Z(\alpha)$ in the limit $\alpha\rightarrow\infty$. The vanishing invariant response is understood as, while the input density involves increasingly sharp peaks and correspondingly the cost of perturbations diverges, the output density instead converges uniformly to a Gaussian because of the smoothing from the channel which effectively merges the input peaks.
In contrast to the vanishing invariant response, the mutual information is finite as both $p(y)$ and $p(y|x)$ are Gaussians in the limit so that we recover the simple variance ratio formula, see SM for details.

\paragraph*{Proof of the large information limit.}
Consider a smooth probability density $p(x)$, and a Gaussian channel $p(y|x)=\mathcal{N}(x,\epsilon)(y) $.
By expanding the input density $p(x)$ around the point $x=y$,  we obtain for the output density
$    p(y) =  p(x)\big{|}_y + \frac{\epsilon}{2}   \, \partial^2_x p(x)\big{|}_y$,
where we used the symmetry $\mathcal{N}(x,\epsilon)(y) = \mathcal{N}(y,\epsilon)(x)$ and we are considering only the first order term in the small variance limit.
Similarly we get
$p(y) \left\langle \partial_x \ln p(x) \big{|} y \right\rangle
    = \partial_x p(x)\big{|}_y \, + \frac{\epsilon}{2} \, \partial^3_x p(x)\big{|}_y $.
Then for the invariant response we get
\begin{equation}
    \left\langle \left( \partial_x \ln p(x) \right)^2 \right\rangle  \left( 1 - \gamma_{x\rightarrow y} \right)   
    = \epsilon \left\langle \left( \partial^2_x \ln p(x) \right)^2 \right\rangle ,   
\end{equation}
where we used $p(y)\big{|}_x = p(x) + \frac{\epsilon}{2}   \, \partial^2_x p(x)$ and integrated by parts.
For the mutual information we calculate the first order expansion
$e^{-2I} = \epsilon \,  2 \pi \, e^{1 + 2 \left\langle \ln p(x) \right\rangle } $,
then the first $\epsilon$ order of the conjecture reads
\begin{equation}\label{large information limit}
    \left\langle \left( \partial^2_x \ln p(x) \right)^2 \right \rangle
    \geq 2\pi \, e^ {1+2 \left\langle \ln p(x) \right\rangle  }  \left\langle \left( \partial_x \ln p(x) \right)^2 \right\rangle.
\end{equation}
This inequality can be proven by applying Jensen's inequality to the left hand side and integrating by parts to obtain
\begin{equation}
    \left\langle \left( \partial_x \ln p(x) \right)^2 \right\rangle
    \geq 2\pi e^ {1+2 \left\langle \ln p(x) \right\rangle  }  ,
\end{equation}
which is Stam's isoperimetric inequality \cite{dembo1991information}.
In other words, the large information limit of our conjecture gives a one-dimensional inequality which is less tight than a proven one.

Let us note that the Gaussian assumption for the channel $p(y|x)$ is not essential as long as higher moments of the distribution give higher $\epsilon$ orders, namely when $\lim_{\epsilon\rightarrow 0}\frac{1}{\epsilon}\left\langle (y-x)^{2+k}\right\rangle = 0$ for $k > 0$.

\paragraph*{Discussion.}
The invariant response was introduced in \cite{auconi2021fluctuation} to formalize causation as a context-independent, invariant quantification of the propagation of perturbations. It was obtained by recasting the fluctuation-response theorem in the space of probability distributions.
In this Letter an inequality is conjectured which formalizes the intuitive understanding that the propagation of perturbations is limited by the mutual information that the corresponding variables share in the unperturbed state.
The large information limit of this conjecture is here proven by leveraging Stam's isoperimetric inequality \cite{dembo1991information}, while more general probability distributions have been studied numerically suggesting that the conjecture could be correct at least for some non-trivial class of densities. 
We discuss applications to the thermodynamics of feedback control \cite{sagawa2010generalized} and to a problem in estimation theory \cite{barnes2021fisher}.
An extension to the three-dimensional case would be needed to properly discuss causation as its minimal framework requires an additional time-lagged variable \cite{auconi2021fluctuation}.

\paragraph*{Acknowledgements.}
I thank Andrea Giansanti, Marco Scazzocchio and Guido Caldarelli for helpful discussions.

\bibliography{apssamp}

\end{document}


\preprint{APS/123-QED}

\title{Supplementary Materials for the manuscript "Conjecture of an information-response inequality"}

\author{Andrea Auconi}
\affiliation{%
 Ca’ Foscari University of Venice, DSMN, via Torino 155, 30172 Mestre (Venice), Italy
}%



\maketitle



\section{Notation convention}\label{Notation convention}

Let us clarify the angle bracket notation \cite{auconi2021fluctuation} with the example functional $\left\langle \left\langle \partial_x \ln p(x) \big{|} y \right\rangle \big{|} x \right\rangle$.
Consider first the internal average $\left\langle \partial_x \ln p(x) \big{|} y \right\rangle \equiv \int dx\, p(x|y) \, \partial_x \ln p(x)$, that is simply the expectation of $\partial_x \ln p(x)$ conditional to $y$. We note that $\left\langle \partial_x \ln p(x) \big{|} y \right\rangle$ is a function of just $y$ since $x$ has been averaged out.
Now we take the average of $\left\langle \partial_x \ln p(x) \big{|} y \right\rangle$ conditional to $x$, namely
\begin{equation}
    \left\langle \left\langle \partial_x \ln p(x) \big{|} y \right\rangle \big{|} x \right\rangle \equiv
    \int dy \, p(y|x) \int dx' \, p(x'|y) \, \partial_{x'} \ln p(x')   ,
\end{equation}
where we have to use the dummy $x'$ to differentiate the statistics $p(x'|y) \equiv p(x|y)\big{|}_{x',y}$ and $p(x') \equiv p(x)\big{|}_{x'}$ in the internal average from the condition $x$ in the external average.
We note that $\left\langle \left\langle \partial_x \ln p(x) \big{|} y \right\rangle \big{|} x \right\rangle$ is a function of just $x$ since $x'$ and $y$ have been averaged out.

\section{Fluctuation-response theorem for Kullback-Leibler divergences.}
Here we rewrite, for the simpler input-output setting considered in this manuscript, the fluctuation-response theorem for Kullback-Leibler (KL) divergences introduced in \cite{auconi2021fluctuation}.

Given an arbitrary smooth channel $p(y|x)$ and an $\epsilon$-translation of the input density, $q(x)=p(x-\epsilon)$, we expand the resulting perturbed output density as
\begin{multline}
    q(y) \equiv \int dx \, q(x) \, p(y|x) \\
    = \int dx \, \left[ p(x) - \epsilon \, \partial_x p(x) +\frac{\epsilon^2}{2} \,\partial^2_x p(x)\right] \, p(y|x)  + \mathcal{O}(\epsilon^3) \\
    = p(y) \left[ 1 -\epsilon \int dx \, p(x|y) \, \partial_x \ln p(x) +\frac{\epsilon^2}{2}  \int dx \, p(x|y) \left( \partial^2_x \ln p(x)  +\left(  \partial_x \ln p(x) \right)^2 \right)
    \right] + \mathcal{O}(\epsilon^3) \\
    = p(y) \left[ 1 -\epsilon \left\langle \partial_x \ln p(x) \big{|} y \right\rangle  +\frac{\epsilon^2}{2}  \left\langle  \partial^2_x \ln p(x)  +\left(  \partial_x \ln p(x) \right)^2  \Big{|} y \right\rangle
    \right] + \mathcal{O}(\epsilon^3) ,
\end{multline}
where in the last line we used the notation convention, and by $\mathcal{O}(\epsilon^3)$ we just mean terms that are negligible compared to $\epsilon^2$ in the limit $\epsilon\rightarrow 0$. Please note that brackets are always meant with respect to the unperturbed densities $p$. For the response divergence we have
\begin{multline}\label{response divergence}
    D\left[ q(y) || p(y) \right] \, \equiv \int dy \, q(y) \, \ln \left( \frac{q(y)}{p(y)} \right) \\
    = \left\langle \left( 1 -\epsilon \left\langle \partial_x \ln p(x) \big{|} y \right\rangle \right) \, \ln \left( \frac{q(y)}{p(y)} \right)  \right\rangle + \mathcal{O}(\epsilon^3) \\
    = \frac{\epsilon^2}{2} \, \left\langle \left\langle \partial_x \ln p(x) \big{|} y \right\rangle ^2  \right\rangle + \mathcal{O}(\epsilon^3),
\end{multline}
where we expanded the logarithm, $\ln(1+\alpha)=\alpha-\frac{\alpha^2}{2} + \mathcal{O}(\alpha^3)$, used iterated conditioning to evaluate $\left\langle \left\langle \partial_x \ln p(x) \big{|} y \right\rangle  \right\rangle = \left\langle \partial_x \ln p(x)  \right\rangle =0 $, and integrated by parts $\left\langle  \partial^2_x \ln p(x)  \right\rangle = - \left\langle \left(  \partial_x \ln p(x) \right)^2 \right\rangle$.
Similarly, for the perturbation divergence we have the standard relation with the Fisher information \cite{amari2016information}, 
\begin{equation}\label{perturbation divergence}
    D\left[ q(x) || p(x) \right] = \frac{\epsilon^2}{2}\left\langle \left( \partial_x \ln p(x) \right)^2 \right\rangle + \mathcal{O}(\epsilon^3),
\end{equation}
where we recall $q(x) = p(x-\epsilon)$.
From Eq. \eqref{response divergence}-\eqref{perturbation divergence} we can write
\begin{equation}
    D\left[ q(y) || p(y) \right] \,= \, \gamma_{x\rightarrow y} \, D\left[ q(x) || p(x) \right] \, + \mathcal{O}\left( (D\left[ q(x) || p(x) \right])^{\frac{3}{2}} \right ),
\end{equation}
that is the fluctuation-response theorem in the space of probability distributions with the KL divergence as a metric. The linear response coefficient
\begin{equation}
    \gamma_{x\rightarrow y} \equiv \lim_{\epsilon\rightarrow 0} \frac{D\left[ q(y) || p(y) \right]}{D\left[ q(x) || p(x) \right]} = \frac{\left\langle \left\langle \partial_x \ln p(x) \big{|} y \right\rangle ^2 \right\rangle}{\left\langle \left( \partial_x \ln p(x) \right)^2 \right\rangle}  ,
\end{equation}
contrary to classical linear response coefficients is invariant under invertible transformations as indeed it is a ratio of KL divergences which are invariant functionals, and we therefore name it \textit{invariant response}.

\section{Data processing inequality.}
Here we give the detailed derivation of the De Bruijn-like identity which enables an equivalent formulation of the invariant response as a data processing ratio.

Consider a small Gaussian noise $z$ with zero mean and $\epsilon$ variance, $p(z)=\mathcal{N}(0,\epsilon)(z)$, and imagine to pass it through the channel $p(y|x)$. More explicitly, we define $\widetilde{x}\equiv x+z$ and consider the mutual information ratio $I(z,\widetilde{y})/I(z,\widetilde{x})$ corresponding to the channel $p(\widetilde{y}|\widetilde{x})\equiv p(y|x)\big{|}_{\widetilde{y},\widetilde{x}}$.
We calculate
\begin{multline}
    p(\widetilde{y}|z) \equiv \int d\widetilde{x} \, p(\widetilde{x}|z)\, p(y|x)\big{|}_{\widetilde{y},\widetilde{x}} \,
    = \int dx \, p(x-z)\, p(y|x)\big{|}_{\widetilde{y},x} \\
    = p(y)\big{|}_{\widetilde{y}} \left[ 1 -z \left\langle \partial_x \ln p(x) \big{|} y \right\rangle\big{|}_{\widetilde{y}}  +\frac{z^2}{2}  \left\langle  \partial^2_x \ln p(x)  +\left(  \partial_x \ln p(x) \right)^2  \Big{|} y \right\rangle\big{|}_{\widetilde{y}}
    \right] + \mathcal{O}(z^3),
\end{multline}
and
\begin{equation}
    p(\widetilde{y})  = p(y)\big{|}_{\widetilde{y}} \left[ 1  +\frac{\epsilon}{2}  \left\langle  \partial^2_x \ln p(x)  +\left(  \partial_x \ln p(x) \right)^2  \Big{|} y \right\rangle\big{|}_{\widetilde{y}}
    \right] + \mathcal{O}(\epsilon^2). 
\end{equation}
Then for the mutual information $I(z,\widetilde{y})$ we find
\begin{multline}
    I(z,\widetilde{y}) \equiv \int \int dz \, d\widetilde{y} \, p(z,\widetilde{y}) \ln\left(\frac{p(\widetilde{y}|z)}{p(\widetilde{y})}\right)\\
    = \int \int dz \, d\widetilde{y} \, p(z) \, p(y)\big{|}_{\widetilde{y}} \left[ 1 -z \left\langle \partial_x \ln p(x) \big{|} y \right\rangle\big{|}_{\widetilde{y}}  +\frac{z^2}{2}  \left\langle  \partial^2_x \ln p(x)  +\left(  \partial_x \ln p(x) \right)^2  \Big{|} y \right\rangle\big{|}_{\widetilde{y}}
    \right] \times \\
    \times    \left[ -z \left\langle \partial_x \ln p(x) \big{|} y \right\rangle\big{|}_{\widetilde{y}}  +\frac{1}{2} \left(z^2-\epsilon\right)  \left\langle  \partial^2_x \ln p(x)  +\left(  \partial_x \ln p(x) \right)^2  \Big{|} y \right\rangle\big{|}_{\widetilde{y}} 
    -\frac{1}{2} z^2 \left(  \left\langle \partial_x \ln p(x) \big{|} y \right\rangle\big{|}_{\widetilde{y}}  \right)^2
    \right]
    + \mathcal{O}(\epsilon^2) \\
    = \frac{\epsilon}{2} \left\langle \left\langle \partial_x \ln p(x) \big{|} y \right\rangle ^2 \right\rangle  + \mathcal{O}(\epsilon^2).
\end{multline}
Then deriving with respect to the variance $\epsilon$ we find
\begin{equation}
    \frac{d}{d\epsilon} I(z,\widetilde{y})\Big{|}_{\epsilon=0} = \frac{1}{2} \left\langle \left\langle \partial_x \ln p(x) \big{|} y \right\rangle ^2 \right\rangle,
\end{equation}
which is a type of De Bruijn's identity \cite{dembo1991information}.
Similarly we find $ \frac{d}{d\epsilon} I(z,\widetilde{x})\Big{|}_{\epsilon=0} = \frac{1}{2} \left\langle \left( \partial_x \ln p(x) \right)^2 \right\rangle$, so that for the invariant response
\begin{equation}
    \gamma_{x\rightarrow y} = \lim_{\epsilon \rightarrow 0} \frac{I(z,\widetilde{y})}{I(z,\widetilde{x})}.
\end{equation}
According to the data processing inequality \cite{dembo1991information,rioul2010information} the channel cannot increase information, $I(z,\widetilde{y}) \leq I(z,\widetilde{x})$, then for the invariant response we have $\gamma_{x\rightarrow y} \leq 1$. Then we see that our conjecture can be interpreted as a particular refined data processing inequality based on the channel mutual information $I(x,y)=I(\widetilde{x},\widetilde{y})\big{|}_{\epsilon=0}$.

\section{Proof of the large information limit.}
Here we give a more detailed derivation of the large information limit to complement the main text.
Consider a smooth probability density $p(x)$, and a Gaussian channel $p(y|x)=\mathcal{N}(x,\epsilon)(y) $, where  $\mathcal{N}(x,\epsilon)(y) \equiv \left( 2 \pi \epsilon \right)^{-1/2}  \exp \left[ - (y-x)^2 / (2 \epsilon )\right]$.
By expanding the density $p(x)$ around the point $x=y$, in the limit of vanishing variance $\epsilon \rightarrow 0$ we obtain
\begin{multline}
    p(y) = \int dx \, p(x) \, p(y|x) \\
    = \int dx \, \mathcal{N}(y,\epsilon)(x) \, \left[ p(x)\big{|}_y  \, +(x-y) \,  \partial_x p(x) \big{|}_y  \, +\frac{1}{2} (x-y)^2 \, \partial^2_x p(x)\big{|}_y \, + \mathcal{O}\left((x-y)^3\right) \right] \\
    = p(x)\big{|}_y + \frac{\epsilon}{2}   \, \partial^2_x p(x)\big{|}_y \, + \mathcal{O} ( \epsilon^2 ) ,
\end{multline}
where we used the symmetry $\mathcal{N}(x,\epsilon)(y) = \mathcal{N}(y,\epsilon)(x)$.
Similarly we have
\begin{multline}
    \left\langle \partial_x \ln p(x) \big{|} y \right\rangle
    \equiv \int dx \, p(x|y) \, \partial_x \ln p(x) 
    = \frac{1}{p(y)} \int dx \, p(y|x) \, \partial_x p(x) \\
    = \frac{1}{p(y)} \int dx \, \mathcal{N}(y,\epsilon)(x) \, \left[ \partial_x p(x)\big{|}_y  \, +(x-y) \,  \partial^2_x p(x) \big{|}_y  \, +\frac{1}{2} (x-y)^2 \, \partial^3_x p(x)\big{|}_y \, + \mathcal{O}\left((x-y)^3\right) \right] \\
    = \frac{1}{p(y)} \left[ \partial_x p(x)\big{|}_y \, + \frac{\epsilon}{2} \, \partial^3_x p(x)\big{|}_y \right] + \mathcal{O} ( \epsilon^2 ),
\end{multline}
and for the square
\begin{equation}
        \left\langle \partial_x \ln p(x) \big{|} y \right\rangle ^2
    = \frac{1}{\left[ p(y) \right]^2}  \left[ \left( \partial_x p(x)\big{|}_y \right)^2 \, + \epsilon \, \partial_x p(x)\big{|}_y \, \partial^3_x p(x)\big{|}_y \right] + \mathcal{O} ( \epsilon^2 ).
\end{equation}
Then for the response Fisher information we get
\begin{multline}
    \left\langle \left\langle \partial_x \ln p(x) \big{|} y \right\rangle ^2  \right\rangle 
    = \int dy \, \frac{1}{p(y)} \left[ \left( \partial_x p(x)\big{|}_y \right)^2 \, + \epsilon \, \partial_x p(x)\big{|}_y \, \partial^3_x p(x)\big{|}_y \right] + \mathcal{O} ( \epsilon^2 ) \\
    = \int dx \, \frac{1}{p(y)\big{|}_x} \left[ \left( \partial_x p(x) \right)^2 \, + \epsilon \, (\partial_x p(x)) \, \partial^3_x p(x) \right] + \mathcal{O} ( \epsilon^2 ) \\
    = \int dx \, p(x)  \left(\partial_x \ln p(x)\right)^2  \left[ 1 - \frac{\epsilon}{2}   \, \frac{\partial^2_x p(x)}{p(x)} \right]
    \,\, + \epsilon \int dx \, \left(\partial_x \ln p(x)\right) \partial_x ^3 p(x) \,
    + \mathcal{O} ( \epsilon^2 )\\
    = \int dx \, p(x)  \left(\partial_x \ln p(x)\right)^2  -\epsilon \int dx \, p(x)  \left(\partial^2_x \ln p(x)\right)^2 \,
    + \mathcal{O} ( \epsilon^2 )\\
    \equiv \left\langle \left( \partial_x \ln p(x) \right)^2 \right\rangle -\epsilon \left\langle \left( \partial^2_x \ln p(x) \right)^2 \right\rangle \,
    + \mathcal{O} ( \epsilon^2 )  .    
\end{multline}
where we used $p(y)\big{|}_x = p(x) + \frac{\epsilon}{2}   \, \partial^2_x p(x) + \mathcal{O} ( \epsilon^2 )$, and then integrated by parts.

For the mutual information we simply have
\begin{equation}
    I \equiv \left\langle \ln \left( \frac{p(x,y)}{p(x) p(y)} \right) \right\rangle = \left\langle \ln p(y|x) \right\rangle - \left\langle \ln p(y) \right\rangle
    = -\frac{1}{2} \ln \left( 2\pi e \epsilon \right) - \left\langle \ln p(x) \right\rangle + \mathcal{O} (\epsilon).    
\end{equation}

The first $\epsilon$ order in the conjecture is then
\begin{equation}\label{large information limit}
    \left\langle \left( \partial^2_x \ln p(x) \right)^2 \right \rangle
    \geq 2\pi e^ {1+2 \left\langle \ln p(x) \right\rangle  }  \left\langle \left( \partial_x \ln p(x) \right)^2 \right\rangle.
\end{equation}

This inequality of Eq. \eqref{large information limit}, which we obtained as a large information limit of our conjecture, can be proven by applying Jensen's inequality to the left hand side, $\left\langle \left( \partial^2_x \ln p(x) \right)^2 \right \rangle
    \geq \left\langle \partial^2_x \ln p(x)  \right \rangle ^2 = \left\langle \left( \partial_x \ln p(x) \right)^2 \right \rangle ^2$, to obtain
\begin{equation}
    \left\langle \left( \partial_x \ln p(x) \right)^2 \right\rangle
    \geq 2\pi e^ {1+2 \left\langle \ln p(x) \right\rangle  }  ,
\end{equation}
which Stam's isoperimetric inequality \cite{dembo1991information}.
In other words, the large information limit of our conjecture gives a one-dimensional inequality which is less tight than a proven one.

We note that the Gaussian assumption for the channel is not essential as long as the scaling is such that higher moments of the distribution give higher $\epsilon$ orders, namely $\lim_{\epsilon\rightarrow 0}\frac{1}{\epsilon}\left\langle (y-x)^{2+k}\right\rangle = 0$ for $k > 0$.

\section{Constrained optimization of the invariant response.}
Here we apply the Lagrange multipliers method to show that the Gaussian family is an extremum in the optimization of the invariant response subject to mutual information and probability normalization constraints,
\begin{equation}
    \max_{p(x,y)} \gamma_{x\rightarrow y} \,\,\,\,\,\,\,\, \Bigg{|}
    \,\,\,\,\,\,\,\,
    \left\langle \ln \left( \frac{p(x,y)}{p(x)p(y)} \right) \right\rangle = I ,
    \,\,\,\,\,\,\,\,
    \left\langle 1 \right\rangle = 1.
\end{equation}
where the invariant response functional $\gamma_{x\rightarrow y} \equiv \gamma_{x\rightarrow y}[p]$ is
\begin{equation}
    \gamma_{x\rightarrow y} = \frac{\left\langle \left\langle \partial_x \ln p(x) \big{|} y \right\rangle ^2 \right\rangle}{\left\langle \left( \partial_x \ln p(x) \right)^2 \right\rangle}  .
\end{equation}
   
The functional derivative of $\gamma_{x\rightarrow y}$ with respect to the density $p(x,y)$ is
\begin{multline}
    \frac{\delta \gamma_{x\rightarrow y} }{\delta p} \,\,\,\,
    = \, \frac{1}{\left\langle \left( \partial_x \ln p(x) \right)^2 \right\rangle} \, \frac{\delta}{\delta p} \left\langle \left\langle \partial_x \ln p(x) \big{|} y \right\rangle ^2 \right\rangle
    \,\,- \frac{\left\langle \left\langle \partial_x \ln p(x) \big{|} y \right\rangle ^2 \right\rangle}{\left\langle \left( \partial_x \ln p(x) \right)^2 \right\rangle ^2 } \, \frac{\delta}{\delta p} \left\langle \left( \partial_x \ln p(x) \right)^2 \right\rangle \\
    = \, \frac{1}{\left\langle \left( \partial_x \ln p(x) \right)^2 \right\rangle} \, 
    \left[  
    - \left\langle \partial_x \ln p(x) \big{|} y \right\rangle ^2
    + 2 \left( \partial_x \ln p(x) \right) \left( \left\langle \partial_x \ln p(x) \big{|} y \right\rangle - \left\langle \left\langle \partial_x \ln p(x) \big{|} y \right\rangle \big{|} x \right\rangle \right)
    - 2 \partial_x \left\langle \left\langle \partial_x \ln p(x) \big{|} y \right\rangle \big{|} x \right\rangle
    \right] \\
    + \frac{\gamma_{x\rightarrow y}}{\left\langle \left( \partial_x \ln p(x) \right)^2 \right\rangle } \, \left[ 2 \partial^2_x \ln p(x) + \left( \partial_x \ln p(x) \right)^2 \right],
\end{multline}
where we computed the functional derivative $\frac{\delta}{\delta p} \left\langle \left\langle \partial_x \ln p(x) \big{|} y \right\rangle ^2 \right\rangle$ as usual by introducing the arbitrary function $\phi(x,y)$ and evaluating the infinitesimal variation
\begin{multline}
    \frac{d}{d\epsilon} \left\langle \left\langle \partial_x \ln p(x) \big{|} y \right\rangle ^2 \right\rangle \left[ p(x,y) + \epsilon \phi(x,y) \right] \Bigg{|}_{\epsilon = 0}\\
    = \frac{d}{d\epsilon} \int dy \left( \int dx' \left[ p(x',y) + \epsilon \phi(x',y) \right]  \right)^{-1} \left( \int dx \left[ p(x,y) + \epsilon \phi(x,y) \right] \partial_x \ln \int dy'  \left[ p(x,y') + \epsilon \phi(x,y') \right] \right)^2 \Bigg{|}_{\epsilon = 0}\\
    = \int \int dx\, dy\, \phi(x,y) \,\left\langle \partial_x \ln p(x) \big{|} y \right\rangle \left( 2  \partial_x \ln p(x) -\left\langle \partial_x \ln p(x) \big{|} y \right\rangle \right) \,
    +2 \int dx \, p(x) \left\langle \left\langle \partial_x \ln p(x) \big{|} y \right\rangle \big{|} x \right\rangle \, \partial_x \left( \frac{\int dy \phi(x,y)}{p(x)} \right)\\
    = \int \int dx\, dy\, \phi(x,y) \, \left[- \left\langle \partial_x \ln p(x) \big{|} y \right\rangle ^2
    + 2 \left( \partial_x \ln p(x) \right) \left( \left\langle \partial_x \ln p(x) \big{|} y \right\rangle - \left\langle \left\langle \partial_x \ln p(x) \big{|} y \right\rangle \big{|} x \right\rangle \right)
    - 2 \partial_x \left\langle \left\langle \partial_x \ln p(x) \big{|} y \right\rangle \big{|} x \right\rangle
    \right] ,
\end{multline}
where we used integration by parts.

The functional derivative of the mutual information is
\begin{equation}
    \frac{\delta}{\delta p} \left\langle \ln \left( \frac{p(x,y)}{p(x)p(y)} \right) \right\rangle = \ln \left( \frac{p(x,y)}{p(x)p(y)} \right) -1,
\end{equation}
and of the normalization is $\frac{\delta}{\delta p} \left\langle 1 \right\rangle = 1$.

Let us write the Lagrangian function
\begin{equation}
    L[p] \equiv  \gamma_{x\rightarrow y} [p]
    - \alpha \left( \left\langle \ln \left( \frac{p(x,y)}{p(x)p(y)} \right) \right\rangle [p] -I \right)
    - \beta \left(\left\langle 1 \right\rangle [p] -1 \right),
\end{equation}
where $\alpha$ and $\beta$ are Lagrange multipliers, and we highlighted with the argument $[p]$ the functionals entering the optimization. In particular, note that $I$ is a constraint and not a functional here.
By imposing the stationarity $\frac{\delta L[p]}{\delta p}=0$ we find
\begin{multline}\label{stationarity}
    \left\langle \left( \partial_x \ln p(x) \right)^2 \right\rangle \left( \alpha \left[ \ln \left( \frac{p(x,y)}{p(x)p(y)} \right) -1 \right]
    +\beta \right) \,
    = \,   
    \gamma_{x\rightarrow y}  \left[ 2 \partial^2_x \ln p(x) + \left( \partial_x \ln p(x) \right)^2 \right] \\
    - \left\langle \partial_x \ln p(x) \big{|} y \right\rangle ^2
    + 2 \left( \partial_x \ln p(x) \right) \left( \left\langle \partial_x \ln p(x) \big{|} y \right\rangle - \left\langle \left\langle \partial_x \ln p(x) \big{|} y \right\rangle \big{|} x \right\rangle \right)
    - 2 \partial_x \left\langle \left\langle \partial_x \ln p(x) \big{|} y \right\rangle \big{|} x \right\rangle  .
\end{multline}
By averaging Eq. \eqref{stationarity} with the constraint $\left\langle \ln \left( \frac{p(x,y)}{p(x)p(y)} \right) \right\rangle = I$ we get
\begin{equation}\label{beta}
    \beta = - \alpha (I-1) ,
\end{equation}
where we integrated by parts $   \left\langle \left( \partial_x \ln p(x) \right) \left\langle \left\langle \partial_x \ln p(x) \big{|} y \right\rangle \big{|} x \right\rangle  \right\rangle =
- \left\langle \partial_x \left\langle \left\langle \partial_x \ln p(x) \big{|} y \right\rangle \big{|} x \right\rangle  \right\rangle $,
and noticed that \\
$  \left\langle \left( \partial_x \ln p(x) \right) \left\langle \partial_x \ln p(x) \big{|} y \right\rangle \right\rangle
  = \left\langle \left\langle \partial_x \ln p(x) \big{|} y \right\rangle ^2 \right\rangle$ .
Then from Eq. \eqref{stationarity}-\eqref{beta} we obtain the functional form
\begin{multline}\label{functional form extremum}
    p(x,y) \,= \, p(x) \, p(y) \, e^I \,
    \exp \left[
    \frac{\gamma_{x\rightarrow y} \left[ 2 \partial^2_x \ln p(x) + \left( \partial_x \ln p(x) \right)^2 \right] }{\alpha \left\langle \left( \partial_x \ln p(x) \right)^2 \right\rangle }
    \right] \, \times \\
    \times \,
    \exp \left[ \frac{- \left\langle \partial_x \ln p(x) \big{|} y \right\rangle ^2
    + 2 \left( \partial_x \ln p(x) \right) \left( \left\langle \partial_x \ln p(x) \big{|} y \right\rangle - \left\langle \left\langle \partial_x \ln p(x) \big{|} y \right\rangle \big{|} x \right\rangle \right)
    - 2 \partial_x \left\langle \left\langle \partial_x \ln p(x) \big{|} y \right\rangle \big{|} x \right\rangle}{\alpha \left\langle \left( \partial_x \ln p(x) \right)^2 \right\rangle} \, 
    \right] ,
\end{multline}
where $\alpha$ is determined by the normalization $\langle 1 \rangle = 1$.

Consider now the bivariate Gaussian family
\begin{equation}\label{Gaussian family}
    p(x,y) = \frac{1}{2\pi \sigma_x \sigma_y \sqrt{1-\rho^2}} \exp \left[ - \,\frac{z}{2 (1-\rho^2)} \right],
\end{equation}
with
\begin{equation}
    z = \frac{\left( x-\mu_x \right)^2}{\sigma^2_x} + \frac{\left( y-\mu_y \right)^2}{\sigma^2_y} 
    - \frac{2\rho (x-\mu_x) (y-\mu_y)}{\sigma_x \sigma_y} ,
\end{equation}
where $\rho$ is the correlation coefficient, and $(\mu,\sigma^2)$ are the mean and variance parameters.
Let us compute the expectations entering the extremum functional form of Eq. \eqref{functional form extremum} with respect to the Gaussian density of Eq. \eqref{Gaussian family}. First we evaluate $\partial_x \ln p(x) = \frac{\mu_x-x}{\sigma^2_x}$, and $\partial^2_x \ln p(x) = -\frac{1}{\sigma^2_x}$, then the expectations
\begin{equation}
    \left\langle \left( \partial_x \ln p(x) \right)^2 \right\rangle = \frac{1}{\sigma_x^2} ,\,\,\,\,\,\,\,\,
    \left\langle \partial_x \ln p(x) \big{|} y \right\rangle = \rho \, \frac{(\mu_y -y)}{\sigma_x \sigma_y},\,\,\,\,\,\,
    \left\langle \left\langle \partial_x \ln p(x) \big{|} y \right\rangle \big{|} x \right\rangle = \frac{\rho ^2}{\sigma_x^2} (\mu_x -x),
\end{equation}
so that we get $\gamma_{x\rightarrow y} = \rho ^2$, and $I=-\frac{1}{2} \ln\left( 1-\rho ^2 \right)$. Therefore for Gaussian systems it holds the equality
\begin{equation}
   \gamma_{x\rightarrow y} = 1 - e^{-2I} .
\end{equation}

Using the above relations we find that by taking $\alpha= 2 (1-\rho^2)$ the Gaussian family satisfies the functional form (Eq. \eqref{functional form extremum}) of the extremum point in the optimization problem. Clearly this does not guarantee the Gaussian family to be a global maximum of the optimization problem and the conjecture to be correct.

\section{Example of vanishing invariant response with finite mutual information.}
Consider the input density
\begin{equation}
    p(x) = \frac{1}{Z(\alpha )}  \, e^{-x^2} \, \cos^2(\alpha x) ,
\end{equation}
where $\alpha$ controls the oscillations frequency and $Z(\alpha ) \, = \int dx \, e^{-x^2} \, \cos^2(\alpha x) = \frac{\sqrt{\pi}}{2} \left( 1 + e^{-\alpha^2} \right)$ is the normalization factor.

Consider as channel the linear Gaussian input-output relation $p(y|x)=\mathcal{N}(x,1)(y)\equiv (2\pi)^{-1/2} \exp\left[ - (y-x)^2 / 2 \right]$. 
The output density is then
\begin{equation}\label{p_y example alpha}
    p(y) \equiv \int dx \, p(x) \, p(y|x) \,
    = \frac{e^{-\frac{y^2}{3}}}{\sqrt{3\pi}} \, \frac{1+e^{-\frac{2}{3}\alpha^2} \cos \left(\frac{2}{3} \alpha y \right)}{1+e^{-\alpha^2}} ,
\end{equation}
where we used the integral $\int dx \,  \cos^2 (\alpha x) \, e^{-\frac{3}{2}x^2 +xy} \,= \sqrt{\frac{\pi}{6}} \, e^{\frac{y^2}{6}} \left[ 1 + e^{-\frac{2}{3}\alpha^2} \cos \left( \frac{2}{3} \alpha y \right) \right]$.
Then we have
\begin{equation}
    p(x|y) \equiv \frac{p(x)\, p(y|x)}{p(y)} \,
    =  \frac{\sqrt{6/\pi} \, e^{-\frac{y^2}{6}}}{1+e^{-\frac{2}{3}\alpha^2} \cos \left(\frac{2}{3} \alpha y \right)}  \, e^{-\frac{3}{2}x^2 +xy} \, \cos^2 (\alpha x) .
\end{equation}
We proceed to evaluate the conditional expectation
$\left \langle \partial_x \ln p(x) \big{|} y \right\rangle \,= -2 \alpha \left \langle \tan (\alpha x) \big{|} y \right\rangle -2 \left \langle x \big{|} y \right\rangle$. The first term is calculated as
\begin{equation}
     \alpha \left \langle \tan (\alpha x) \big{|} y \right\rangle \,
    =  \alpha \, e^{-\frac{2}{3}\alpha^2} \frac{\sin \left( \frac{2}{3} \alpha y \right)}{1 + e^{-\frac{2}{3}\alpha^2} \cos \left( \frac{2}{3} \alpha y \right)},
\end{equation}
where we used the integral $\int dx \,  \sin (\alpha x) \, \cos (\alpha x) \, e^{-\frac{3}{2}x^2 +xy} \,= \sqrt{\frac{\pi}{6}} \, e^{\frac{y^2}{6} -\frac{2}{3}\alpha^2} \sin \left( \frac{2}{3} \alpha y \right)$.
For every value of $y$ we can upper bound the absolute value of this contribution by
\begin{equation}
    \Big{|} \alpha \left \langle \tan (\alpha x) \big{|} y \right\rangle \Big{|}  \leq
     \frac{\alpha \, e^{-\frac{2}{3}\alpha^2}}{1 - e^{-\frac{2}{3}\alpha^2}},
\end{equation}
which vanishes for $\alpha\rightarrow \infty$. Therefore $\alpha \left \langle \tan (\alpha x) \big{|} y \right\rangle$ converges uniformly to zero and we can neglect it with respect to the other term $\left \langle x \big{|} y \right\rangle$ in the limit, so that
\begin{equation}
    \lim_{\alpha \rightarrow \infty} \left \langle \left \langle \partial_x \ln p(x) \big{|} y \right\rangle^2 \right\rangle  \,
    = 4 \lim_{\alpha \rightarrow \infty} \left \langle \left \langle x \big{|} y \right\rangle^2 \right\rangle \,
    \leq  4 \lim_{\alpha \rightarrow \infty} \left \langle \left \langle x ^2 \big{|} y \right\rangle \right\rangle \,
    = 4 \lim_{\alpha \rightarrow \infty}  \left \langle x^2 \right\rangle = 2 < \infty,
\end{equation}
where we used Jensen's inequality, iterated conditioning, and evaluated the integral limit $\,\lim_{\alpha \rightarrow \infty} \int dx\, x^2 \, e^{-x^2} \, \cos^2(\alpha x) = \sqrt{\pi} / 4$ by using the trigonometric identity $\cos^2 (\alpha x) = \frac{1}{2} \left( 1 + \cos (2 \alpha x)   \right)$, the Gaussian integral $\int dx\, x^2 \, e^{-x^2} = \sqrt{\pi} / 2 $, and the Riemann–Lebesgue Lemma to limit the integral $\,\lim_{\alpha \rightarrow \infty} \int dx \, x^2 \, e^{-x^2} \, \cos (2 \alpha x) = 0$. 
So we showed that the response Fisher information is bounded in the limit, $\lim_{\alpha \rightarrow \infty} \left \langle \left \langle \partial_x \ln p(x) \big{|} y \right\rangle^2 \right\rangle < \infty$.

Similar to the above, we evaluate the fast oscillations limit of the perturbation Fisher information and obtain the asymptotic divergence $\,\lim_{\alpha \rightarrow \infty} \frac{1}{\alpha^2} \left \langle  \left( \partial_x \ln p(x)  \right)^2  \right\rangle = 2\sqrt{\pi}$. 

This qualitative difference between perturbation and response Fisher information limits is understood as, while the input density involves increasingly sharp peaks and correspondingly its Fisher information diverges, the output density instead converges uniformly to a Gaussian because of the smoothing from the channel which effectively merges the input peaks, as it can be seen from Eq. \eqref{p_y example alpha}, so that the response Fisher information converges to its Gaussian counterpart and is therefore finite. With this we showed that the invariant response vanishes in the limit of fast oscillations, $\,\lim_{\alpha \rightarrow \infty} \gamma_{x\rightarrow y} = 0$.

From the probability densities and expectations calculated above we can immediately evaluate the fast oscillations limit of the mutual information $I(\alpha) \equiv \left\langle \ln \left(\frac{p(x,y)}{p(x)p(y)} \right)  \right\rangle$. We get
$\lim_{\alpha \rightarrow \infty} I(\alpha) = \frac{1}{2} \ln\left(\frac{3}{2}\right)$,
which is a finite number. This is understood as both $p(y)$ and $p(y|x)$ converge uniformly to Gaussians in the limit so that we recover the simple variance ratio formula. Also note that $p(x|y)$ is still oscillatory in the limit being based on the prior $p(x)$.

\section{Numerical evidence.}
We studied the conjecture numerically on a set of random distributions of the form
\begin{equation}\label{random densities}
    p(x,y) = \frac{K^2(x,y)}{Z} \exp \left[ - \left(\frac{x^2+y^2}{2 \alpha}\right)^\beta \right] ,
\end{equation}
where $K(x,y)=k_{00}+k_{10}x+k_{01}y+\frac{1}{2}k_{20}x^2+k_{11}xy + ... + k_{03}y^3 $ is a 3rd-order Taylor polynomial around $(0,0)$, the functional $Z$ ensures normalization, and the parameters $\alpha$ and $\beta$ control the nonlinearity. We sampled in the parameter space of $K(x,y)$ and for several $\alpha$ and $\beta$ configurations, and we did not yet observe violations of the conjecture.


\bibliography{apssamp}